\documentclass[a4paper,eqsecnum,aps,showpacs,floatfix,byrevtex,twocolumn]{revtex4}
\usepackage{graphicx}
\usepackage{amsfonts}
\begin{document}
\title{Magnetic properties of $\alpha$-iron(II) phthalocyanine}
\author{M. Evangelisti}
\affiliation{Kamerlingh Onnes Laboratory, Leiden University, 2300 RA, Leiden, The Netherlands,}
\affiliation{Instituto de Ciencia de Materiales de Arag\'on, C.S.I.C. - Universidad de Zaragoza,
50009 Zaragoza, Spain}
\author{J. Bartolom\'e}
\affiliation{Instituto de Ciencia de Materiales de Arag\'on, C.S.I.C. - Universidad de Zaragoza,
50009 Zaragoza, Spain}
\author{L. J. de Jongh}
\affiliation{Kamerlingh Onnes Laboratory, Leiden University, 2300 RA, Leiden, The Netherlands}
\author{G. Filoti}
\affiliation{National Institute for Materials Physics, P.O. Box MG-07, 76900 Bucharest-Magurele,
Romania}
\date{\today}
\begin{abstract}
We report on the magnetic properties of the supra-molecular compound iron(II) phthalocyanine in its
$\alpha$-form. dc- and ac-susceptometry measurements and M\"ossbauer experiments show that the iron
atoms are strongly magnetically coupled into ferromagnetic Ising chains with very weak
antiferromagnetic interchain coupling. The transition to 3D magnetic ordering below 10~K is
hindered by the presence of impurities or other defects, by which the domain-wall arrangements
along individual chains become gradually blocked/frozen, leading to a disordered 3D distribution of
ferromagnetic chain segments. Below 5~K, field-cooled and zero-field-cooled magnetization
measurements show strong irreversible behavior, attributed to pinning of the domain-walls by the
randomly distributed defects in combination with the interchain coupling. High-field magnetization
experiments reveal a canted arrangement of the moments in adjacent ferromagnetic chains.
\end{abstract}


\maketitle

\section{Introduction}

\label{intro}Metal phthalocyanines (hereafter referred to as MPc) have received extensive attention
in the last three decades because of their commercial applications as dyes, catalysts, and because
of their interesting electro-optical properties~\cite{book,cook}. Their magnetic properties have
attracted attention as well since manganese phthalocyanine (MnPc) was one of the first molecular
magnets~\cite{mitra}. In this work we focus on iron(II) phthalocyanine (FePc).

In the MPc molecule, the M atom has square planar coordination with four pyrrolic N atoms. In the
bulk and in thin films, these planar molecules form linear chains by columnar stacking. The
molecules are tilted by an angle with respect to the chain axis~\cite{barra,honig}. The angle at
the parallel adjacent chain is opposite, thus constructing a so-called ``herringbone'' structure.
The different angles and relative positions of the chains give rise to different polymorphs, for
example in the $\alpha$ and $\beta$ forms of copper phthalocyanine (CuPc)~\cite{honig}. This is of
importance since the magnetic properties of metal complexes are strongly dependent on structural
differences. For example, the particular structure in the $\beta$-MnPc is thought to underly the
Mn--Mn ferromagnetic interactions that make this compound the only ferromagnetic MPc compound so
far reported~\cite{mitra}.

A very similar molecule with square planar M coordination is M-porphyrine. These molecules may also
stack in different polytypes, giving rise to different magnetic properties. Very recently it was
reported that Fe(II) octaethyltetraazaporphyrin crystallizes in two structures: the $\alpha$-form
which is a canted, soft molecular ferromagnet with $T_{C}=5.6$~K, and the $\beta$-form which
remains paramagnetic down to 1~K~\cite{yee,sellers}. Apparently, though the Fe(II) electronic
states are identical, the different exchange paths and distances between the neighboring molecules
and between chains modulate the magnetic behavior. Thus, also Fe(II) square coordinated complexes
seem prone to show various types of magnetic behavior.

FePc could fall into this category of magnetic complexes. There are two forms of FePc described in
the literature. The $\alpha$-form is metastable and obtained either as a polycrystalline powder or
as a thin film formed by vacuum deposition onto a cold substrate~\cite{harri,ercolani}. The
$\beta$-form is the most stable one and is obtained not only from the sublimation technique in the
form of a single crystal, but also from heating the $\alpha$-form sample to 350$^{\circ}$~C for a
few hours~\cite{harri}. Since 1935, the $\beta$-form of this compound has been repeatedly reported
to behave as a singlet-ground-state paramagnet down to the lowest measured
temperatures~\cite{barra,klemm,lever,dale}. On the other hand, to the best of our knowledge, the
magnetic properties of $\alpha$-FePc have so far never been studied. We show below that in the
$\alpha$-form of FePc the Fe(II) moments are strongly coupled into ferromagnetic chains, with the
weak interchain coupling leading to a a canted, soft molecular ferromagnet below about 10~K. Our
findings demonstrate that FePc has a behavior very similar to MnPc, so the magnetic behavior of the
latter is not so exotic as thought earlier~\cite{mitra}, but is a member of a more general category
of square-planar metal complexes in which ferromagnetic correlations are established. They
represent, therefore, quite interesting examples of molecular magnets, a subject of intensive
on-going research~\cite{res}.

\section{Experimental details}\label{expe}

All the experiments were performed on powdered samples of the $\alpha$-FePc compound as obtained
from Aldrich company (catalogue number 37,954.9), in two different batches. We verified the
stoichiometry by elemental analysis. The morphology was studied by scanning electron microscopy
with a resolution of 3.5~nm. A step scanned powder X-ray diffraction pattern was collected at room
temperature using a rotating anode X-ray diffractometer. The diffractometer was operated at 80~mA
and 40~kV with Cu radiation and a graphite monochromator was used to select the Cu~K$\alpha_{1,2}$
radiation. Data were collected from $4^{\circ}$ up to $90^{\circ}$ with a step size of
$0.02^{\circ}$ and a counting rate of 4~s/step.

Magnetic data were obtained with the use of a commercial SQUID based ac-magnetometer instrument
down to 1.7~K with an external magnetic field in the range $-50$ to 50~kOe. The amplitude of the
ac-field was 4.5~Oe in all the experiments and the frequency $f$ was varied between 0.1 and
1380~Hz. Magnetic data with larger applied field (up to 200~kOe) were obtained in the Nijmegen High
Field Magnet Laboratory. The magnetization was measured using the ballistic extraction technique
and the sensitivity was about $10^{-3}$~emu. The heat capacity was measured from 1.8~K up to 20~K
with a commercial calorimeter.

\section{Morphology and structure}

\label{ms} By scanning electron microscopy, we have observed that $\alpha$-FePc develops in
crystallites whose shape is similar to a `bunch of rods' (Fig.~1). The ratio of length to thickness
is approximately 10. Without any preferred length, the crystallites can be as long as up to
hundreds of micrometers. By means of powder X-ray diffraction we have found that the present form
of FePc is isomorphic to the $\alpha$I-form of 4-monochloro-copper phthalocyanine~\cite{honig},
that is triclinic with space group $P1$ or $P\bar{1}$. The fit of the X-ray diffraction spectrum
provides the following cell dimensions: $a=12.2$~\AA, $b=3.78$~\AA, $c=13.0$~\AA,
$\alpha=90.0^{\circ}$, $\beta=90.4^{\circ}$ and $\gamma=95.6^{\circ}$. The fit to the peak
intensities is not good due to the needle shape of the powder grains that produces a texture
effect. Thus, in the present sample of $\alpha$-FePc the molecules are stacked parallel to each
other along the short $b$-axis at intervals of 3.78~\AA, and the molecular planes are inclined to
the $ac$ plane at an angle of 26.5$^{\circ}$, in a herringbone structure (Fig.~2). The vertical
distance between the neighboring molecules is about 3.4~\AA. We have verified further that the
present form of FePc is indeed the (metastable) $\alpha$-form because it shows the characteristic
IR-absorption peak at 250~cm$^{-1}$~\cite{ercolani}. By warming our sample at 350$^{\circ}$~C
during 24 hours in vacuum, the compound transforms into the $\beta$-form, as expected.

In the $\alpha$-form, the Fe atom is coordinated to the four N atoms in the diagonal positions of
the square planar molecule. The nearest neighbors on the adjacent molecule at one side are the two
isoindole and one azamethine nitrogens located at similar distances, while at the other side the
coordination is identical but rotated by 180$^{\circ}$. The Fe(II) ion is in a symmetry lower than
$D_{4h}$, but in a first approximation we shall assume this local symmetry. The value of
3.78~\AA~for the Fe--Fe separation along the $b$-axis in the present form of FePc is very
reasonable, lying within the accepted range of 3.5--4.8~\AA~for metalloporphyrin
dimers~\cite{boyd}. It is generally accepted that the Fe--Fe separation along the $b$-axis and the
angle of inclination of the molecular planes with respect to the crystallographic $ac$ plane
determine the extent of the exchange Fe--Fe interaction.

We note that the main structural difference with the $\beta$-form consists in the inclination angle
(Fig.~2); in the $\beta$-form it is 45.8$^{\circ}$ and the azemitine nitrogen of each adjacent
molecule lays just on the apical position of an octahedron of N atoms~\cite{honig}, in the
$\alpha$-form it is 26.5$^{\circ}$, and the N atoms from the nearest molecules are not in axial
positions and therefore do not form such a N octahedron about the Fe atom.

\section{Experimental results}\label{mp}

The inverse of the in-phase component of the zero-field ac-susceptibility of $\alpha$-FePc, as
measured at $f=90$~Hz, is plotted as a function of temperature in Fig.~3. As shown by the solid
line, the data can be fitted between 50~K and 300~K by a Curie-Weiss law $\chi=C/(T-\theta)$ for
$S=1$, $g=2.54$ and $\theta=(40\pm 2)$~K. This result implies that in this temperature range the
iron(II) ion is in an effective spin $S=1$ state, and the large positive value of $\theta$
indicates that there are strong ferromagnetic correlations between the Fe moments in the
paramagnetic phase. On basis of the powder X-ray structural analysis, we expect these correlations
to extend within the columnar chains along the $b$-axis. Neglecting interchain magnetic coupling
and using the mean-field expression for the Curie-Weiss temperature $\theta=2zJS(S+1)/3k_{{\rm
B}}$, we find for $S=1$ and $z=2$ the estimate $J/k_{{\rm B}}\approx 15$~K for the ferromagnetic
interaction strength along the chains. We note that also for the $\beta$-form of FePc, an effective
spin $S=1$ was found below room temperature. Indeed, at high temperatures, the susceptibilities of
the $\alpha$ and $\beta$~\cite{barra,labarta} forms are coincident (inset of Fig.~3). The fact that
the susceptibility of the $\beta$-form goes to zero for $T\rightarrow 0$ has been explained in
terms of zero-field splitting of the triplet state, leaving a non-magnetic singlet as the lowest
energy level occupied for $T\le 20$~K~\cite{barra}.

Turning now to the data taken below 50~K, as exemplified in Fig.~4, we note that the behavior of
the susceptibility of $\alpha$-FePc points to strong low-dimensional magnetic
characteristics~\cite{miedema}. Indeed, as seen in Fig.~4, the susceptibility continues to increase
down to temperatures much lower than $\theta$. Only at about 10~K the ac-susceptibility curve shows
a change of slope, and a rounded maximum is found at still lower temperatures. The paramagnetic
behavior is thus sustained down to approximately 10~K, that is much less than $\theta$, as expected
for a magnetic chain system~\cite{miedema}.

From the shape of the $\chi\prime(T)$ curve in Fig.~4, one would be tempted to associate the change
of slope to the onset of 3D correlations between the chains due to antiferromagnetic interchain
coupling. However, here we should point to the fact that $\chi^{\prime}$ at the maximum is not very
far below the calculated limit for a completely ferromagnetically ordered isotropic material below
$T_{C}$, namely $1/N$ with $N$ the demagnetizing factor (a rough estimate gives the value of
$\approx 33$~emu/mol).

Since the presence of anisotropy can lower the value of the susceptibility of a ferromagnetic
powder considerably~\cite{miedema,anis}, the occurrence of ferromagnetic interchain ordering would
also be considered. Moreover, in case the moments of the ferromagnetic chains would be subject to
strong crystal field anisotropy, the latter could compete with the interchain interactions, and,
when sufficiently strong, lead to a freezing at some blocking temperature before the interchain
coupling becomes effective (analogous to the phenomena observed in molecular magnetic clusters and
superparamagnetic particles).

In order to further investigate the nature of the ``ordering'' processes, we have therefore
measured the zero-field differential susceptibility in the region below 15~K for different
frequencies. The dependence on frequency of the real and imaginary parts of the ac-susceptibility
are shown in Fig.~5. As was already seen for $f=90$~Hz in Fig.~4, for all frequencies used the
in-phase component of the ac-su\-sce\-pti\-bi\-li\-ty, $\chi^{\prime}$, shows a rounded maximum in
the range 5~K--10~K. This is always accompanied by a maximum in the out-of-phase component,
$\chi^{\prime\prime}$, at somewhat lower temperature. The maximum in $\chi^{\prime}(T)$ shifts to
lower temperature with decreasing frequency, and $\chi^{\prime}(T)$ decreases abruptly to zero
below the maximum. The sharp maximum in $\chi^{\prime\prime}(T)$ shifts likewise with frequency. It
is not possible then to define a definite transition temperature to a 3D ordered state from these
data, since the obtained $T_{C}$ would be frequency dependent. The shift in temperature of the
maxima in $\chi^{\prime}(T)$ and $\chi^{\prime\prime}(T)$ with the frequency are reminiscent of
spin-glass or superparamagnetic behavior. Since the freezing temperature $T_{f}(\omega)$ can be
defined from $\omega\tau[T_{f}(\omega), H=0]=1$, a quantitative measure of the frequency shift is
obtained from $(\Delta T_{f}(\omega)/T_{f}(\omega))/\Delta {\rm ln}~(\omega)=0.034$. It is six
times larger than the rate for a metallic spin glass and an order of magnitude smaller than for a
superparamagnet~\cite{spinglass}.

In view of the just-described frequency dependence of the ac-susceptibility, we decided to
investigate the analogy with spin glass systems or superparamagnets still further, by means of
dc-magnetization studies. In Fig.~6 we show dc-$M(T)$ measurements taken with the SQUID
magnetometer in the zero-field cooling (ZFC) and field-cooling (FC) modes at two fields, $H=50$~Oe
and 1~kOe. We note that both curves show a change in slope at about 10~K, similar as for
$\chi^{\prime}$ in Figs.~4 and 5. Furthermore, a bifurcation point showing strong irreversibility
is found below about 5~K. At first sight, the bifurcation temperature is not much affected by the
field although we point out that for the low-field curve (50~Oe) the ZFC and FC curves strictly
speaking only coincide above $\simeq 10$~K.

In another experiment, after a FC process down to 1.8~K with $H=40$~Oe the field was switched off
and the remnant magnetization was measured (Fig.~7). We found a large remnant magnetization at 1.8
K which very rapidly decays by several orders of magnitude when increasing the temperature towards
5~K. Above 5~K, the decay of the remanence proceeds much slower (inset of Fig.~7), and a small
amount of remnant magnetization is seen to persist even up to 10~K--15~K. On basis of Fig.~7 we may
conclude that, although most of the field-induced magnetization is indeed lost at about 5~K, a
small fraction of about 3\% is left above this temperature, and is only removed by warming up to
about 15~K. This behavior points to the presence of substantial ferromagnetic short-range order
along the chains in this intermediate region of 5~K--15~K, in agreement with the behavior of the
ac-susceptibility.

The $M(H)$ measurements of the initial (ZFC) magnetization were therefore extended to higher fields
and at temperatures above, at and below this intermediate region in order to get some more insight
in these irreversibility processes (Fig.~8). The data at $T=30$~K show the expected paramagnetic
behavior. In the intermediate temperature region, for $T=8$~K and 5~K, the initial increase in
$M(H)$ is very abrupt, reaching the value $1.4~\mu_{B}$ for $H=2.5$~kOe. Above 10~kOe, the
magnetization curves measured with the SQUID up to 50~kOe (Fig.~8) or in the Nijmegen High Field
Magnet Laboratory (Fig.~9), show an additional slow increase up to $\sigma=1.7~\mu_{B}$ at
$H=200$~kOe. Note in Fig.~9 the perfect coincidence of the SQUID equilibrium measurement with the
branch corresponding to the high-field measurement at decreasing field, while the branch measured
at increasing field is lower. The SQUID measurements have a characteristic point-to-point measuring
time of $\sim 150$~s while the high-field measurements have a characteristic time of $\sim 40$~s.
The system evidently needs a longer time than 40~s to reach equilibrium at this temperature when
the field is increased from zero.

The spontaneous magnetization evaluated as the extrapolation to $H=0$ of the high field
demagnetizing curve yields a value of $1.5~\mu_{B}$. We point out that, although these measurements
demonstrate that in the region below about 5~K the compound is in a ferromagnetic state, when the
$g$ factor deduced in the paramagnetic region is still valid in this range, the saturation
magnetization would be $2.5~\mu_{B}$ on basis of the (effective) spin $S=1$ ground state found for
the Fe(II) ions from the Curie-Weiss analysis of the high-temperature susceptibility (Fig.~3). We
shall come back to this discrepancy below in the discussion section, where we shall argue that on
basis of the competition between the exchange term and the crystal field term in the Hamiltonian
for $\alpha$-FePc, a substantial canting of the magnetic moments with respect to the crystal field
(anisotropy) axis is to be expected. The increases seen in the $M(H)$ curves above 5~kOe may then
be associated with the gradual rotation of the canted moments, produced by the competition between
the Zeeman energy and the crystal field energy.

One may conclude that below 5~K, the material shows strong irreversibility, as already evidenced by
the bifurcation point in Fig.~6. Also the behavior of the $M(H)$ curves in Fig.~8 points to the
presence of long relaxation times. For example, the initial $M(H)$ curve measured at 1.8~K is
actually even lower than the data at 15~K, since the time waited to measure each point is of the
order of minutes and only a fraction of the sample is oriented. This is more clearly seen in the
hysteresis loop measured at 1.8~K by sweeping the field between $-50$ and 50~kOe in a time of $\sim
150$~s (Fig.~10). The initial curve increases progressively with the field, tending to a value of
$1.6~\mu_{B}$ at 20~kOe. When the field decreases, the magnetization remains nearly constant down
to zero-field, yielding a remanence of $1.5~\mu_{B}$. However, the inversion of the remnant moment
is very abrupt when a field in the opposite direction is applied, indicating that there is no
coercivity in the usual sense associated with this remanence.

Similarly strong relaxation effects in the low-temperature range have been seen in preliminary
$^{57}$Fe M\"ossbauer spectroscopic studies on $\alpha$-FePc. In Fig.~11 spectra at some
representative temperatures are shown. Above 25~K, the spectrum is hardly temperature dependent and
reveals the familiar paramagnetic, quadrupolar split doublet. Below that temperature the doublet
starts gradually to broaden and below about 15~K a magnetic hyperfine splitting appears, at first
extremely broadened, then becoming the sharper the lower the temperature. Below about 5~K, the
broadening has almost disappeared and a fully hyperfine split spectrum is observed. A detailed
presentation of the M\"ossbauer spectra will be given in another publication.

Similar temperature dependent relaxation phenomena as seen in the M\"ossbauer spectra in Fig.~11,
have typically been seen for blocking phenomena in superparamagnetic particles or for domain-wall
excitations in anisotropic magnetic (Ising) chains. Briefly, the magnetic moment of the M\"ossbauer
nucleus (here $^{57}$Fe) feels the hyperfine field exerted on it by the electron spin on the same
atom. Thus the fluctuations in the hyperfine field are directly related to the electron spin
dynamics, i.e. the nuclear moment probes the autocorrelation function of the electron spin.
Accordingly, the M\"ossbauer spectrum will be broadened as long as the fluctuation rate of the
electron spin lies in the frequency window ($10^{6}$~Hz--$10^{9}$~Hz) of the M\"ossbauer
experiment. The temperature dependence observed indicates that the electron spin dynamics is
described by thermally activated processes. At high temperatures the fluctuation rate of the
electron spins is too fast to cause broadening. With lowering temperature the rate slows down,
enters the M\"ossbauer frequency window leading to the strong relaxation. At the lower temperatures
the rate becomes just too slow to be felt. In superparamagnetic particles, the activated process is
related to the thermally excited switching of the particle's magnetic moment between the
orientations dictated by the anisotropy, which becomes blocked at low temperatures. In the magnetic
Ising chain systems, the fluctuation rate is dictated by the density of thermally excited
domain-walls (solitons) along the chains. At high temperatures, the density is very high and thus
the fluctuations too fast. At intermediate temperatures, the fluctuation rate passes the
M\"ossbauer window, whereas, at low temperatures, the soliton density becomes too small, moreover
the walls may become blocked/pinned by interchain interactions or defects in the chains.

\section{Discussion}\label{disc}

Summing up the evidence obtained so far, we may state that the magnetic structure of $\alpha$-FePc
is most probably composed of very weakly coupled ferromagnetic chains, with an intrachain exchange
$J/k_{{\rm B}}\approx 15$~K, and an effective spin $S=1$ for the Fe(II) ion in the temperature
range 50~K--300~K. The crystal field splitting, responsible for the effective spin $S=1$, likely
result in addition into a strong anisotropy of the ferromagnetic interaction. Below, we shall thus
start the discussion with a full analysis of the crystal field effects in this material, in
particular also in relation to the other members of the metal phthalocyanine and metal porphyrine
family. Next we shall present the ensuing analysis of the magnetic susceptibility in the range
10~K--50~K in terms of 1D magnetic models systems. Finally, the nature of the relaxation and
ordering phenomena observed below 15~K will be further analyzed.

As is well known, for Fe(II) the combination of orthorhombic crystal field splitting combined with
spin-orbit coupling leads to an effective spin description for the five lowest occupied levels
$(d_{xy},d_{yz},d_{xz},d_{z^{2}},d_{x^{2}-y^{2}})$~\cite{app1,app2}. Depending on the ratio of the
parameters, and the relevant temperatures of the experiment, one can have a situation where either
two, three or all five levels will participate in the magnetic ordering processes, leading to an
effective spin $S=1/2$, 1 or 2, respectively, for the Fe(II) ion. For thermal energies in the range
100~K--300~K, the $\alpha$-FePc may be described as an effective spin $S=1$ with $g=2.54$. However,
the thermally occupied levels, which lead to the $S=1$, are split into a lower-lying singlet and an
excited doublet, with the splitting being probably of the order of 50~K. Accordingly, the magnetic
behavior will be determined by the competition between the magnetic exchange energy and the crystal
field term representing this singlet-doublet splitting. From the Curie-Weiss fit to the
susceptibility, we have seen that $\theta\approx 40$~K, so both these terms are of the same order
of magnitude. Below we will show that the competition may actually produce a canted arrangement of
the ferromagnetically aligned spins, with an effective spin $S=1/2$ at low temperatures and
considerable anisotropy.

For the temperature range 100~K--300~K, we conclude that the Fe(II) ion in $\alpha$-FePc may be
described as an effective spin $S=1$ compound, in which single-ion anisotropy and second-order
spin-orbit coupling leaves the triplet as the lowest energy multiplet. This is the same spin state
as found in $\beta$-FePc~\cite{barra}. Indeed, both compounds have identical magnetic properties at
room temperature, which arise from a similarly split triplet [$D/k_{{\rm B}}\simeq 53$~K and 98~K,
for $\alpha$ and $\beta$-form, respectively (see below)] with the non-magnetic singlet as the
ground state~\cite{labarta}.

In the $\alpha$-form the $S=1$ spins remain ferromagnetically coupled in magnetic chains down to
the lowest temperatures in spite of the non-magnetic ground singlet since the ferromagnetic
interaction $2zJ$ (with $J/k_{{\rm B}}\approx 15$~K as estimated from the Curie-Weiss fit) is
larger than $D$, in agreement with Moriya's criterion for magnetic ordering in a $S=1$
triplet~\cite{moriya}. Apparently, this condition is not satisfied in $\beta$-FePc because the
ration of $D$ over $J$ is larger, and as a consequence it remains paramagnetic at all temperatures.

Curiously, the magnetic phenomenology of the MnPc forms is just opposite: i.e. the $\beta$-form is
ferromagnetic while the $\alpha$-form is antiferromagnetic~\cite{yamada}. In the Mn compounds the
different behavior of the two forms was attributed to competing ferro- and antiferromagnetic
exchange interactions, neglecting direct Mn--Mn coupling. Under the assumption that the electronic
configuration of the $3d^{5}$ electrons is $(d_{xy})^{2}(d_{xz})^{1}(d_{yz})^{1}\-(d_{z^{2}})^{1}$,
the ferromagnetic interaction is caused by the overlap between the $d_{z^{2}}$ unpaired electron
($a_{1g}$ symmetry) with the $\pi$ orbital of $E_{g}$ symmetry of the adjacent aromatic ring, which
is orthogonal to the $d_{z^{2}}$ orbital of its metal ion (Fig.~12). The antiferromagnetic coupling
is due to the overlap of the $d_{xz}d_{yz}$ orbitals ($e_{g}$) with the same $\pi$ orbital of
$E_{g}$ symmetry (Fig.~12). In the MnPc $\beta$-form the ferromagnetic mechanism dominates while in
the MnPc $\alpha$-form the ferromagnetic path weakens and antiferromagnetism
prevails~\cite{yamada}.

If we assume that the electronic configuration of Fe is related to that of the Mn by just adding
one electron to the second level as proposed for the $\beta$-form of FePc by Dale et
al.~\cite{dale} and Coopens et al.~\cite{copp}, that is $E_{g}A$
$(d_{xy})^{2}(d_{xz})^{2}(d_{yz})^{1}(d_{z^{2}})^{1}$, then the same arguments as for the Mn
compounds would be applicable. The only difference would be that the antiferromagnetic interaction
should be weakened, since the $(d_{xz},d_{yz})$ doublet should have just one unpaired electron.
Consequently, the $\beta$-form should be ferromagnetic, contrary to observation. One arrives to the
same conclusion if one assumes the electronic configuration $B_{2g}$
$(d_{xy})^{1}(d_{xz})^{2}(d_{yz})^{2}(d_{z^{2}})^{1}$, as proposed by Barraclough et
al.~\cite{barra}.

The conclusion is that the electronic configuration needs to be different, or direct exchange is
more intense in the $\alpha$-form than in the $\beta$-form. It can be argued that the orbital
ground state $A_{2g}A$ $(d_{xy})^{2}(d_{xz})^{1}(d_{yz})^{1}(d_{z^{2}})^{1}$, proposed by Stillman
et al.~\cite{still} leads to just antiferromagnetic coupling. The remaining configuration $E_{g}B$
$(d_{z^{2}})^{2}(d_{xz})^{2}(d_{yz})^{1}(d_{xy})^{1}$, proposed by Labarta~\cite{labarta},
generates a path for ferromagnetism in the $\alpha$-form: the unpaired electron in the $d_{xy}$
orbital singlet may interact with its counterpart in the adjacent molecule via the $E_{g}$ $\pi$
electron of the aromatic ring, which is orthogonal, giving rise to its ferromagnetic character
(Fig.~12). This is the only ferromagnetic exchange path possible, however its intensity would be
probably very weak since it is in the $xy$ plane and the overlap will be small. At any rate, in the
$\beta$-form the $d_{xy}$ orbital has negligible overlap with the $E_{g}$ electron of the apical N
since it is located in the $z$ axis, thus the ferromagnetic interaction would be negligible. The
unpaired electron of the $(d_{xz},d_{yz})$ doublet would lead in both forms to antiferromagnetic
interaction. So, exchange interaction is not likely to cause the ferromagnetic coupling present in
the $\alpha$-form.

On the other hand the condition for an enhancement with respect to the $\beta$-form of the direct
Fe--Fe interaction are fulfilled since the Fe--Fe distance is reduced from 4.8~\AA ($\beta$) to
3.8~\AA ($\alpha$). Following Goodenough~\cite{good}, the direct ferromagnetic interaction occurs
when there is overlap between a half filled and an empty orbital, or a half-filled and a full
orbital. Since the $(d_{x^{2}-y^{2}})$ empty orbital is very high in energy~\cite{labarta} we think
that only the case of half-filled full-orbital overlap may be relevant here. Then, if the orbital
configuration is $E_{g}A$ $(d_{xy})^{2}(d_{xz})^{2}(d_{yz})^{1}(d_{z^{2}})^{1}$, as proposed by
Coppens et al. for the $\beta$-form from accurate X-ray diffraction~\cite{copp}, the overlap may
take place between two $(d_{xz})^{2}(d_{yz})^{1}$ orbitals. The proposition of the $E_{g}A$
electronic configuration by these authors is based on very direct observation of electronic
deformation densities and in turn supports the proposed mechanism for the ferromagnetism in the
$\alpha$-form. Of course, this overlap will decay strongly when the atoms are separated further by
1~\AA, and so the ferromagnetic interaction weakens to the extent of being negligible in the
$\beta$-form.

Thus, we conclude that the difference in the phenomenology of the Mn and Fe forms can be related to
the different ground state spin configuration (two magnetic doublets for Mn, and one magnetic
doublet and a non-magnetic singlet for Fe) on the one hand, and to the different electron filling
scheme and overlap probabilities of the resulting orbitals, on the other.

As said in the introduction, such a different behavior, ferromagnetism or paramagnetism, between
two herringbone structured compounds formed by stacked square planar molecules containing Fe, has
also been detected in Fe(II) octaethyltetraazaporphyrin~\cite{yee,sellers}. Similarwise, the
$\alpha$-form is ferromagnetic while the $\beta$-form is paramagnetic. However, in a recent
paper~\cite{reiff} a magnetic study of Mn(II) octaethyltetraazaporphyrin shows that the
$\alpha$-form is also ferromagnetic while the $\beta$-form is paramagnetic, in contrast to the
behavior of the MnPc forms. So, the parallelism between the phthalocyanines and
octaethyltetraazaporphyrin is not complete.

On the basis of the above, it appears to be beyond doubt that the magnetic moment of the Fe(II) in
the $\alpha$-FePc is to be associated with a spin triplet probably split by crystal field into a
doublet and a lower-lying singlet. We then have to reconcile this conclusion with the rather low
values of the magnetization reached even in high fields as exemplified in Figs.~8--10. Based on an
effective spin $S=1$ and the powder $g$-value of 2.54 deduced from the Curie-Weiss fit to the
susceptibility above 50~K, the saturation moment should amount to 2.5~$\mu_{{\rm B}}$ per ion, i.e.
far above the measured value even for an applied field of 200~kOe. Looking more closely at the
measured $M(H)$ curves and neglecting for the moment the relaxation effects, one may observe that
below about 10~K the magnetization process can be divided in two steps. From the $M(H)$ curve
measured at 5~K (Fig.~8), a very rapid increase of the magnetization up to 1.0~$\mu_{{\rm B}}$ (per
atom) is observed for fields up to 400~Oe, followed by a much slower gradual increase up to the
highest field applied. This two-step behavior can be explained in terms of a division of the
magnetic chains into two different magnetic sublattices, with a mutually canted arrangement. In
that case the fast rapid increase would correspond to saturation of the moments within each of the
magnetic sublattices, followed by the gradual rotation of the sublattices to a common alignment
parallel to the field. In the latter process, the Zeeman energy has to compete with the (large)
crystal field anisotropy energy, so that the high-field susceptibility (slope of the $M(H)$) curve
can be very small. It is of interest to point out in this regard that for $\beta$-MnPc a canted
magnetic structure in the above sense has indeed been reported~\cite{miyo}.

At first sight the preposition that the ordered magnetic structure is canted seems to contradict
the positive value of $D$, which implies a single crystal anisotropy of planar character. Indeed,
Miyoshi et al.~\cite{miyo} concluded that for chains consisting of ferromagnetic coupled Fe moments
with isotropic exchange, and with single-ion anisotropy, such that the quantization axis of the Fe
sites in one chain is tilted by an angle which is opposite respect to the tilt angle in the
adjacent chain, a negative $D$ is needed to stabilize a canted structure, irrespectively if the
interchain interaction is ferro or antiferromagnetic. This prediction, when applied to the MnPc
compound, revealed a contradiction between the canted structure found in that compound by
magnetization measurements and polarized neutron diffraction~\cite{mitra}, and the positive $D$
determined earlier from the analysis of the high-temperature susceptibility~\cite{barra2}, a
discrepancy that has gone unexplained so far. This apparently identical contradiction, now found
for $\alpha$-FePc and $\beta$-MnPc, can be explained by noting that, contrary to the assumption of
Miyoshi et al.~\cite{miyo} that the ferromagnetic intrachain interaction is isotropic, we find from
our experiments that it is instead of the anisotropic Ising-type, with the easy axis perpendicular
to the molecule plane. Interestingly, the same result; i.e. Ising intrachain interaction, was found
to fit best the MnPc high temperature susceptibility by Barraclough et al.~\cite{barra2}, although
it was not exploited. Thus, the Ising character of the interaction and its larger strength,
comparable to the crystal field term, creates an uniaxial exchange anisotropy that opposes the
crystal-field anisotropy. A rough mean field calculation for FePc, using the value $J/k_{{\rm
B}}\approx 15$~K found from the Curie-Weiss $\theta$, tells us that $H_{ex}=(2zJ/k_{{\rm
B}})\langle S_{z}\rangle/g\mu_{{\rm B}}=58$~T (at $T=0$~K). Such an intense field splits the upper
$S_{z}=\pm 1$ doublet and brings the $S_{z}=-1$ level 30~K down so that it nearly coincides with
the $S_{z}=0$ level. The net effect is that, in terms of the $S=1$ formalism, the magnetic moment
is polarized towards the quantization direction (local $z$ axis) at low temperatures, which is the
necessary condition to have the canted structure according to Miyoshi et al.~\cite{miyo}. In MnPc
the exchange field is $H_{ex}=48$~T ($D/k_{{\rm B}}=29$~K), and as a consequence the $S_{z}=-3/2$
state is pulled below the $S_{z}=-1/2$ state by 68~K (at $T=0$~K) and the ground state gives rise
to an effective uniaxial anisotropy. Thus the annoying contradiction of an apparent inversion of
the $S_{z}=-3/2$ and $S_{z}=-1/2$ states above and below the transition, mentioned at the end of
Ref.~\cite{mitra}, is resolved. In the paper by Reiff et al.~\cite{reiff} on Fe-porphyrine, a
similar qualitative argument of opposing exchange and crystal field effects is mentioned to explain
the magnetic properties of those compounds.

The logical conclusion from the model we propose is that each chain has its own ferromagnetic axis
tilted from the $b$-axis by a given angle, with the easy axis in the adjacent chain tilted by the
opposite angle, for reasons of symmetry. Thus, in a domain formed by this ordered magnetic
structure there is a direction of uncompensated moments, and a perpendicular direction of
compensated moments. The uncompensated moments should have the direction of the $b$-axis in case
the interchain interaction is ferromagnetic, and perpendicular to $b$ if it is antiferromagnetic.
In either cases, there are two degenerate types of domains, in contrast to the four of MnPc, due to
the different symmetry. We refrain from pushing any further the analysis since we have insufficient
experimental information on the magnitude and sign of the interchain interaction nor on the precise
tilt angle, because our measurements are performed on powder samples.

We remark that the above analysis points to two different scenarios for analyzing the magnetic
susceptibility data below 50~K, namely in terms of ferromagnetic Ising chains with either effective
spin $S=1$, or effective spin $S=1/2$, when considering only the lowest two levels ($S_{z}=0$ and
$S_{z}=-1$). Below we shall investigate the applicability of both these possibilities.

Starting with $S=1$, we have calculated the parallel susceptibility of an $S=1$ linear chain under
the assumption of an Ising-like exchange interaction and of an uniaxial single-ion anisotropy $D$
as described by the Hamiltonian:

\begin{displaymath}
{\mathcal
H}_{0}=-2J_{1}\sum_{i=1}^{N-1}S_{z,i}S_{z,i+1}+D\sum_{i=1}^{N}S_{z,i}^{2}
-g\mu_{B}H_{z}\sum_{i=1}^{N}S_{z,i}.
\end{displaymath}

The eigenvalues $\lambda_{\pm}$, $\lambda_{0}$ of this Hamiltonian were obtained by the
conventional transfer matrix technique, similarly as described in Ref.~\cite{brien}, yielding:

\begin{eqnarray}
\lambda_{\pm}=\frac{1}{2}(\gamma +\alpha +1)\pm \frac{1}{2} [(\alpha +\gamma
-1)^{2}+8\beta^{2}]^{1/2}\\ \nonumber\lambda_{0}=\alpha -\gamma,
\end{eqnarray}

where $\alpha ={\rm exp}[(2J_{1}-D)/k_{{\rm B}}T]$, $\gamma ={\rm exp}[(-2J_{1}+D)/k_{{\rm B}}T]$
and $\beta ={\rm exp}(D/2k_{{\rm B}}T)$, with $\lambda_{\pm}$ being the largest eigenvalue.
Utilizing a Taylor series expansion to obtain the eigenvalues in the non-zero-field case, the
susceptibility can be expressed as

\begin{equation}
\label{xsuno} \chi=\frac{Ng^{2}\mu_{{\rm B}}^{2}S(S+1)}{k_{{\rm B}}T} \frac{{\rm e}^{2d}({\rm
e}^{j}\lambda_{+}-1+{\rm e}^{j})}{F_{\lambda}}
\end{equation}

where

\begin{equation}
F_{\lambda}=3\lambda_{+}^{2}-2\lambda_{+}(1+2{\rm e}^{j+2d})-
[2{\rm e}^{2d}(1-{\rm e}^{j})-2{\rm e}^{4d}{\rm sinh}(2j)],
\end{equation}

with $j=2J_{1}/k_{{\rm B}}T$ and $d=-D/2k_{{\rm B}}T$.

Least squares fits of Eq.~(\ref{xsuno}) to the experimental susceptibility data were carried out.
Since the contribution of the perpendicular susceptibility has been neglected, only the data for
relatively high temperatures ($T>10$~K) were used. The dotted line in Fig.~4 shows the best fit
with $J_{1}/k_{{\rm B}}=25.7$~K, $g=2.54$ and $D/k_{{\rm B}}=53.2$~K, where the positive $J_{1}$
indicates ferromagnetic interaction and the positive $D$ implies that the $S_{z}=\pm 1$ doublet
lies above the $S_{z}=0$ state. The results are sensitive to both the temperature range utilized
and the $g$ value assumed. We have tried to fit the same data with the isotropic $S=1$ Heisenberg
model and uniaxial anisotropy~\cite{neef}, but the fit is much worse (Fig.~4). For the temperature
range from 20 to 300~K, the following result was obtained: $J_{1}/k_{{\rm B}}=16.2$~K and
$D/k_{{\rm B}}=76.7$~K. Since we are fitting powder data while the magnetic properties are very
probably highly anisotropic, the values obtained from the fit have to be considered as approximate.
However, we note that the anisotropy constant is about half the value of $D/k_{{\rm B}}=98$~K found
in the $\beta$-form~\cite{dale}. Obviously, the different anisotropy originates from the absence of
the apical N atoms next nearest to the Fe in the $\alpha$-form respect to the $\beta$-form. The
most important result, though, is that the Fe--Fe intrachain interaction is unambiguously
ferromagnetic. The value obtained for $J_{1}$ is in fact similar to that following from the
Curie-Weiss fit.

We next turn to an analysis of the paramagnetic susceptibility below 50~K in terms of the effective
spin $S=1/2$ formalism (Fig.~4), which would apply when either the action of a crystal field of
orthorhombic or lower symmetry, or the combined action of crystal field and magnetic exchange,
would produce an effective doublet as the only occupied energy levels at low temperatures. In that
case, since the effective anisotropy for the doublet will be of the Ising-type, we may compare the
experimental powder susceptibility data with the theoretical predictions for the parallel
($\chi_{\parallel}$) and perpendicular ($\chi_{\perp}$) susceptibilities for the ferromagnetic
$S=1/2$ Ising chain~\cite{katsura}:

\begin{eqnarray}\label{eqkatsura}
\chi_{\parallel}=(Ng_{\parallel}^{2}\mu_{{\rm B}}^{2}/2J_{1/2})K{\rm exp}(2K)\\
\chi_{\perp}=(Ng_{\perp}^{2}\mu_{{\rm B}}^{2}/J_{1/2})[{\rm tanh}(K)+K{\rm cosh}^{-2}(K)]
\end{eqnarray}

where $K=J_{1/2}/2k_{{\rm B}}T$, from which we have calculated the powder susceptibility
$\chi_{p}=(\chi_{\parallel}+2\chi_{\perp})/3$. We can consider also that due to the large expected
anisotropy one has $g_{\parallel}\gg g_{\perp}$, and that the saturation magnetization at low
temperature is $M_{p}=\frac{1}{3}(g_{\parallel}+2g_{\perp})S\mu_{{\rm B}}$, for which the
experiment yielded $1.5\mu_{{\rm B}}$. Fitting both quantities (for $T>10$~K) leads to the
estimates $J_{1/2}/k_{{\rm B}}=32.2$~K, $g_{\parallel}=8.6$ and $g_{\perp}=0.2$. The fit of the
magnetic susceptibility data with the prediction from the above equations is shown in Fig.~4. It is
difficult to judge whether the fit on the basis of the $S=1$ model is better than that with the
$S=1/2$ prediction since: ($a$) the deviations seen for $T\le 10$~K from the $S=1/2$ curve may well
be due to the interchain interactions, which are not taken into account; ($b$) for the same reason
the seemingly better fit with the $S=1$ model (Fig.~4) may be fortuitous since it also neglects
interchain interactions; ($c$) as a further complication the $\chi_{\perp}$ contribution had to be
neglected in the $S=1$ calculation since no theoretical prediction is available.

Proceeding next to a discussion of the ordering/blocking phenomena observed below 10~K, we can
start from the assertion that we may describe $\alpha$-FePc as a system of very weakly coupled
ferromagnetic Ising chains with either effective spin 1/2 or 1. For such systems, it is well-known
that the 1D magnetic behavior is dictated by the excitation of domain-walls (solitons) along the
chains. For strongly anisotropic chains, the width of the domain-wall becomes smaller than a
lattice constant and we speak of ``kinks'' separating spin-up and spin-down regions. At least two
types of excitations in such Ising chains are possible. Namely, the single-kink excitations at the
ends of the chains (with excitation energy $J$) and the kink-pair excitations (with energy $2J$).
The correlation length along the chain  (and thus the magnitude of the magnetic susceptibility) is
determined by the number $n_{k}$ of kinks excited ($\xi\propto n_{k}\propto {\rm exp}~(-J/k_{{\rm
B}}T)$).

Thus we think the present system falls in the same class of Ising-type chains studied previously,
a.o. by magnetic susceptibility and M\"ossbauer effect studies, in connection with the soliton
problem (see e.g. Refs.~\cite{josb,josc,josd,jose,josf,josg,josh,josi}). In fact, the behavior
observed here appears to be analogous to that of FeCl$_{2}$Py$_{2}$, one of the ferromagnetic Ising
chains studied in these previous works. However, for $\alpha$-FePc, the interchain coupling turns
out to be much weaker than in FeCl$_{2}$Py$_{2}$ and related compounds.

As already mentioned when presenting the M\"ossbauer data in Fig.~11, the extremely strong
broadening of the spectra observed in the range 5~K--20~K may indeed be explained in terms of a
spin dynamics based upon the excitation of domain-walls along the ferromagnetic chains. Since the
M\"ossbauer effect is sensitive to high-frequency fluctuations, the relaxation effects already
start to have an influence at about 20~K, whereas relaxation effects in the low-frequency
susceptibility appear only below 10~K. In principle, both the high- and low-frequency spin dynamics
observed can be related to a gradual slowing down of the spin fluctuations with temperature due to
the decrease of the density and the mobility of the walls, followed by (almost) complete pinning
below 5~K. It should be realized that, even when pinned, small excursions around equilibrium
positions of the walls may still remain possible, which explains the possibility of simultaneous
observation of both the high- and low-frequency spin dynamics in the range between 10~K and 5~K.

At this point it is important to recall the similarity of the here-observed relaxation phenomena
with those known for superparamagnetic particles pointed out several times in the above. Indeed,
one might envisage the ferromagnetic chains in $\alpha$-FePc as superparamagnetic particles,
switching their magnetization among the up and down directions along the local anisotropy axis.
However, it should then be immediately realized that the rotation of the spins along a given chain
cannot occur simultaneously, in view of the extremely large anisotropy barrier involved in such a
process (i.e. the anisotropy per spin times the number of spins in a correlated chain). In full
analogy with the situation in ferromagnetic nanowires~\cite{braun}, the rotation has to occur via
the excitation of a small reversed domain, either at the chain end by a $\pi$-wall, or along the
chain by a kink-antikink pair, in both cases followed by the propagation along the chain of the
wall(s). It is clear then, that the mobility of the domain-wall(s) along the chains may become an
important factor in determining the speed of the switching process. In this respect the magnetic
interactions between the chains, which tend to establish full 3D order between correlated chain
segments as the temperature is lowered, can be expected to slow down the wall motions along the
individual chains. It is tempting, therefore, to associate the slowing down of the fluctuations in
the range between 10~K and 5~K, seen in both the M\"ossbauer spectra and the ac-susceptibility, as
a reflection of the gradual establishment of 3D magnetic order between the chains. Since no clear
transition at a definite temperature $T_{C}$ is indicated, the resulting ``order'' is apparently
not long-range but with a substantial amount of randomness, as in a spin glass. Judging from the
behavior seen in e.g. Figs.~6 and 7 of the low-field magnetization, this ordering process is about
completed below 5~K. The source of randomness is probably a random pinning of the domain-walls on
impurities along the chain, whereby the 3D magnetic correlations between the chains are easily
broken up. In view of the instability of $\alpha$-FePc in favor of the $\beta$-FePc form, the
occurrence of such impurities is not unlikely.

The above scenario can also explain the fact that in our specific heat data measured in the range
1.8~K to 20~K, no sign of a singularity (as associated with 3D long-range magnetic ordering) was
detected. The measured specific heat curve just shows a smooth continuous behavior as expected for
a combination of phonon contributions and 1D magnetic fluctuations within the chains.

Adopting therefore tentatively the value of 5~K as representative temperature for the ordering
process between the chains, we may estimate the interchain coupling $J^{\prime}$ either from the
Oguchi's formula~\cite{oguchi}, or from the mean-field argument given by Villain~\cite{villain}, in
both cases obtaining $J^{\prime}/k_{{\rm B}}\approx (10^{-3}-10^{-2})$~K. This estimate may be
confronted with the behavior of the magnetization curves taken at 5~K and 8~K (Fig.~8), in which as
discussed before a rapid initial ``saturation'' mechanism is observed at an applied field of 400~Oe
only. In the canting mechanism discussed above, we have ascribed this low-field process with an
alignment of all the spins on the individual sublattices. This implies the removal by the field of
all domain-walls along the chains, as well as the breaking up of the interchain interactions. It
follows that the effective field $H^{\prime}=z^{\prime}J^{\prime}S/g\mu_{B}$ should be smaller than
400~Oe, which yields (taking $g\simeq 2$, and $S=1$) the upper limit $J^{\prime}/k_{{\rm
B}}\lesssim 0.03$~K. This would suggest in fact that the breaking up of the interchain couplings
occurs in even much smaller fields already and that the magnetization process up to 400~Oe mainly
involves the removal of the domain-walls from the chains by the Zeeman energy.

The quite peculiar form of the hysteresis in Fig.~10 appears to agree with the just sketched
scenario. After first saturating the chain system, the ferromagnetic order within and between the
chains is seen to remain intact until the applied field is reduced to effectively zero, whereafter
the magnetization discontinuously drops to zero. The instability of the ferromagnetic order in
zero-field even at 1.8~K indicates that the interchain interaction is most probably
antiferromagnetic. As soon as the applied field has become smaller than this antiferromagnetic
coupling, it will tend to order the ferromagnetic chains back again into an antiparallel
arrangement. This means that half of the chains will have to switch their magnetization and as we
have argued in the above, this can only occur by the excitation and subsequent propagation of
domain-walls. Part of these walls will then become pinned again so that the antiferromagnetic
arrangement can not become complete and when subsequent by increasing the field in the opposite
direction, the same cycle of removal of the walls has to be overcome.

\section{Concluding remarks}

Clearly at variance with all previously reported behavior of both polytypes of iron(II)
phthalocyanine, we have found that the magnetic structure of $\alpha$-FePc can be described in
terms of ferromagnetic Ising chains, with very weak antiferromagnetic interchain coupling. As a
consequence of impurities and other possible lattice defects, the transition to 3D ordering is
suppressed, so that instead, a gradual freezing/blocking transition is observed to a disordered
magnetic state. The disorder is lifted by a small field of about 2~kOe. Below 5~K this transition
is hindered by strong relaxation effects, which we attribute to the pinning of the domain-walls by
the randomly distributed defects in combination with the interchain coupling. The competition
between magnetic exchange interaction and crystal field effect gives rise to a canted arrangement
of the ferromagnetically aligned spins. Above 10~K the behavior of the susceptibility can be
interpreted in terms of the thermal excitation of kinks (domain-walls) and kink-antikink pairs.
Preliminary M\"ossbauer experiments can be consistently explained in the same terms, the kink
dynamics being reflected in the spectra in the form of strong relaxation effects. The excitation
and dynamics of the kinks lead to line-broadening and hyperfine-splitting of the spectra at
elevated temperature, i.e. far into the paramagnetic region.

\section*{Acknowledgments}
This paper is dedicated to Prof. Dr. Domingo Gonz\'alez Alvarez on the occasion of his retirement.
We are indebted to F. Luis, F. Palacio and J. F. Fern\'andez for helpful discussions, M. C. Sanchez
for the X-ray experiments in Zaragoza, J. A. A. J. Perenboom and H. van Luong for the high-field
magnetization experiments at the FOM-IGM facility in Nijmegen, and M. Laguna for the chemical
analysis in Zaragoza. The financial support inside of Spain--Romania Scientific Co-Operation
program by the Ministerio de Asuntos Exteriores and the Agentia Nationala pentru Stiinta,
Tehnologie si Inovare is gratefully acknowledged. This work was also partially supported by Grant
No. MAT99/1142 from CICYT.

\clearpage

LIST OF FIGURES

\begin{enumerate}
\item{Images of powder $\alpha$-FePc.}
\item{Molecular structure of FePc and its stacking along the $b$-axis.}
\item{Inverse of the in-phase zero-field susceptibility as a function of temperature for an
exciting field with frequency $f=90$~Hz. In the inset: the same data but as $\chi T$ versus $T$.
The susceptibility of $\beta$-FePc is reported for comparison. The solid curve is the fit to the
Curie-Weiss law $\chi=C/(T-\theta)$, where $C=g^{2}\mu_{{\rm B}}^{2}S(S+1)/3k_{{\rm B}}$.}
\item{In-phase zero-field susceptibility as a function of temperature for an exciting field with
frequency $f=90$~Hz. The lines are the calculated susceptibilities in terms of the effective spin
$S=1$ (above) and $S=1/2$ (below). For explanations see text.}
\item{Low-temperature in-phase and out-of-phase susceptibilities for several frequencies.}
\item{Molar magnetization vs temperature measured during heating after ZFC ($\bigtriangledown$) and
FC ($\bigtriangleup$) in an applied field $H_{ap}=50$~Oe and $H_{ap}=1$~kOe.}
\item{Remanence of the magnetization after a FC process with $H=40$~Oe. In the inset: magnification
of the ``lower slope'' region.}
\item{Molar magnetization vs applied field recorded by increasing the field and at different
temperatures: $T=1.8$~K, 5~K, 8~K, 15~K and 30~K. The measuring time is $\approx 150$~s. In the
inset: magnification of the magnetization recorded at 5~K for low-field values.}
\item{High field molar magnetization recorded at $T=5$~K varying the field from null to 200~kOe and
to null again with a measuring time of $\approx 40$~s. For comparison, the data from Fig.~8 are
depicted (filled symbols).}
\item{Hysteresis loop recorded at $T=1.8$~K. A field of 50~kOe is first
applied and then data are collected by sweeping the field to $-50$~kOe and back.}
\item{Representative M\"ossbauer spectra taken at different temperatures. Above 25~K, the shape of
the spectra is almost independent of temperature.}
\item{A schematic of $a_{g}$--$E_{g}$ interaction for $\alpha$-form (a) and $\beta$-form (b), and
$e_{g}$--$E_{g}$ interaction for $\alpha$-form (c) and $\beta$-form (d).}
\end{enumerate}

\end{document}